\documentclass[aps,twocolumn,showpacs,amsmath,amssymb,floatfix]{revtex4-1}

\usepackage{graphicx}
\usepackage{amsmath,bm,amssymb,amsfonts}
\usepackage{lineno,hyperref}
\modulolinenumbers[5]
\begin{document}

\title{Transport properties of Rashba conducting strips
 coupled to magnetic moments with spiral order}

\author{Jos\'e A. Riera}
\address{Instituto de F\'{\i}sica Rosario (CONICET) and
Universidad Nacional de Rosario, Rosario, Argentina}

\date{\today}

\begin{abstract}
Magnetic and transport properties of a conducting layer with Rashba
spin-orbit coupling (RSOC) magnetically coupled to a layer of
localized magnetic moments are studied on strips of varying width.
The localized moments are free to rotate and they acquire an order that
results from the competition between the magnetic exchange energy
and the kinetic energy of the conduction electrons. By minimizing the
total Hamiltonian within the manifold of variational spiral orders
of the magnetic moments, the phase diagram in the space of the
interlayer exchange $J_{sd}$, and the ratio of the Rashba coupling
to the hopping integral, $\lambda/t$ was determined.  Two main phases
with longitudinal spiral order were found, one at large interlayer
coupling $J_{sd}$ with uniform order in the transversal direction,
and the other at small $J_{sd}$ showing a transversal staggered order.
This staggered spiral order is unstable against an antiferromagnetic
(AFM) for large values of $\lambda/t$. In both spiral phases, the 
longitudinal spiral momentum that departs from the expected linear
dependence with the RSOC for large values of $\lambda/t$. Then,
various transport properties, including the longitudinal Drude weight
and the spin Hall conductivity, inside these two phases are computed
in linear response, and their behavior is compared with the ones for
the more well-studied cases of a fixed ferromagnetic (FM) and AFM
localized magnetic orders.

\end{abstract}

\maketitle

\section{Introduction}
\label{introsection}

There is currently an increasing interest in studying and developing
new systems and devices that could process information using the
electron spin, which is the essence of the field of spintronics 
\cite{prinz,wolf,zutic}.
In particular, a considerable number of possibilities stem from the
implementation of effective couplings derived from microscopic
spin-orbit (SO) interactions, chief among them the Rashba spin-orbit 
coupling (RSOC) which appears in systems with structural inversion
asymmetry and leads to the appearance of transversal spin currents
and the spin Hall effect \cite{rashba,hirsch99,sinovaRMP,hoffmann13}.

It has been recently noticed \cite{manchon08,manchon09} that
a strong spin torque can be induced on a two-dimensional (2D) 
conducting layer with Rashba SOC coupled to a ferromagnetic 
(FM) layer. This process was observed when an electrical current
flows in the plane of a Co layer with asymmetric Pt and AlO$_x$
interfaces \cite{miron10,xwang,pesin}. Even more recently,
it has been discussed the possibility of an analogous relativistic
SO torque in conducting layers containing RSOC in the presence of
antiferromagnetic (AFM) layers \cite{sinova14}. This possibility
was suggested that could be realized in bulk Mn$_2$Au, which although
is centrosymmetric, it can be divided into two sublattices that
separately have broken inversion symmetry and form inversion
partners. Various other heterostructures could be considered as 
containing a subsystem of localized magnetic moments with FM or
AFM order \cite{zelezny17}. It has also been noticed that in AFM
coupled systems, the spin-orbit torque could drive magnetic walls
with velocities a magnitude greater than in FM ones 
\cite{gomonay}.

In addition to these FM and AFM orders that are fixed by
large exchange interactions between the localized magnetic moments,
or by the structure of the materials, the case in which the magnetic
moments are allowed to move in order to minimize the total energy,
has also been studied \cite{bode2007,kim2013}. In particular it
has been shown that the spiral order of the
localized magnetic moments along the longitudinal $x$-axis is
driven by the SO interaction in the conduction strip
\cite{kim2013}, which induces a spin rotation or chiral precession,
around the transversal $y$-axis, with $k_{\theta,x} \sim \lambda/t$.
The electron spin spiral in the uncoupled Rashba conducting strip
had been first noticed by Ref.~\cite{datta-das}. Using numerically
exact Monte Carlo calculations within the spin-fermion decoupling,
it has been also shown that in two-dimensional systems, the FM and
AFM orders are unstable against various other types of magnetic
orders, mainly spiral orders \cite{meza}.

Hence, the main motivation for this work is to examine transport 
properties of Rashba conducting strips coupled to a layer of magnetic
moments with the spiral order that minimizes the total energy for each
set of parameters. Spirals, as well as other magnetic orders that may
be present at oxide heterostructures, such as skyrmions, have been
studied but within effective, spin-only, models \cite{xli2014}. In
most of this previous work, infinite two-dimensional systems were
considered and a parabolic band was assumed
\cite{bode2007,kim2013,jia2011}.

It should be noticed that although in conventional semiconductor
heterostructures \cite{nitta} or at the interface LaAlO$_3$/SrTiO$_3$
 \cite{caviglia}, the spin-orbit parameter $\alpha_R$, related to the
coupling $\lambda$ as $\alpha_R=\lambda a$, where $a$ is the lattice
constant, is up to 10$^{−2}~eV~\AA$, a number of compounds where
$\alpha_R$ is more than 2 orders of magnitude higher have been found.
This is the case of BiTeI \cite{sakano,bawden}, in the
BaIrO$_3$/BaTiO$_3$ heterostructure \cite{zzhong}, and in the
CH$_3$NH$_3$PbBr$_3$ organic-inorganic perovskite \cite{niesner}.
In these cases, $\lambda/t \approx 0.3$, which justifies the range
of $\lambda/t$ between 0 and 1 adopted in the present study. It is
also important to emphasize that RSOC can be varied by an electric
field perpendicular to the strip plane, and hence the spiral state could
be accordingly modified. This electric control of magnetic order is at 
the heart of magnetic ferroelectrics or multiferroics \cite{yamasaki}.

Finally, an array of recently proposed materials and devices presents
strong Rashba SO couplings and involve sizable electron fillings
 \cite{bucheli,lmostointer}. Then, in the present work, the coupled 
system with quarter-filled band for the conduction electrons will
be examined for varying Rashba SO interaction and exchange coupling
$J_{sd}$, on strips of various widths between the minimal value
$W=2$ and large enough values to represent the infinite width limit.

The outline of the paper is the following. In Section~\ref{modelmethod}
the model here studied is defined and some methodological details are
provided. Then, in  Section~\ref{phdiag} the phase diagram in the
$\lambda/t$-$J_{sd}$ plane for strips of various widths at quarter
filling is presented. In Section~\ref{helicur}, the behavior of the
Rashba helical currents introduced in Ref.~\cite{hamad}, is discussed, 
and in Section~\ref{perflux}, results for transport properties,
the longitudinal optical conductivity and the spin Hall conductivity
are presented. Finally, in Section~\ref{conclusions}, a summary is 
provided together with a suggestion of possible application of the
present results to spintronic devices.

\section{Model and methods}
\label{modelmethod}

\begin{figure}
\includegraphics[width=0.38\columnwidth,angle=0]{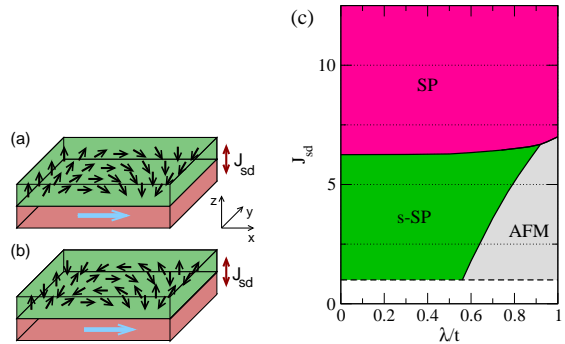}
~~~\includegraphics[width=0.42\columnwidth,angle=0]{phdiagn05.eps}
\caption{(Color online) (a), (b) Schematic depiction of the system
considered in the present work consisting of a layer of localized
magnetic moments coupled by an exchange $J_{sd}$ to a conducting layer
with hopping $t$ and Rashba SO coupling $\lambda$. The magnetic
moments present a spiral ordering in the longitudinal direction that
is uniform (panel (a)) or staggered (panel (b)) in the transversal
or $y$-direction. A charge current along the strip or $x$-direction
that could be injected after a voltage bias is applied to the strip's
ends, is also shown inside the conducting layer.
(c) Schematic phase diagram of model Eq.~(\ref{ham1orb}) on strips
at $n=0.5$. The main phases are the uniform spiral (SP) shown in (a),
the staggered spiral (s-SP) shown in (b), and the antiferromagnetic
(AFM) phase.}
\label{fig1}
\end{figure}

The system to be studied in the present work is schematically
shown in Fig.~\ref{fig1}(a)-(b). A slab of conducting electrons
that are to undergo conventional or spin conserving hopping and
Rashba-type or spin flipping hopping (bottom slab in both pictures)
is coupled by the exchange integral $J_{sd}$ to a slab of 
localized magnetic moments (upper slab). Both slabs are modelled by
a single layer. This heterostructure is similar to the one studied
for the ferro- or antiferromagnetic orders of the magnetic layer
\cite{manchon08,miron10,jia2011}. The crystal structure of the
whole system is assumed cubic, with the layers belonging to the
$x,y$-plane as shown in Fig.~\ref{fig1}(a)-(b).
The Hamiltonian for the resulting model on the $x,y$-plane is
 \cite{sinova14,meza}:
\begin{align}
H_{1o}&=H_{0} + H_{int}   \nonumber  \\
H_{0}&=- t \sum_{<l,m>,\sigma} (c_{l\sigma}^\dagger c_{m\sigma} +
    H. c.) + \lambda \sum_{l}
     [c_{l+x\downarrow}^\dagger c_{l\uparrow}  \nonumber  \\
  &-c_{l+x\uparrow}^\dagger c_{l\downarrow} + i (
   c_{l+y\downarrow}^\dagger c_{l\uparrow}
  + c_{l+y\uparrow}^\dagger c_{l\downarrow}) + H. c.] \nonumber  \\
H_{int}&=-J_{sd} \sum_{l} {\bf S}_l \cdot {\bf s}_l +
         J \sum_{<l,m>} {\bf S}_l \cdot {\bf S}_m
\label{ham1orb}
\end{align}
where $H_{0}$ corresponds to the conducting layer, and it includes the
hopping and the RSOC terms with coupling constants $t$ and $\lambda$,
respectively, which connects nearest neighbor sites on the square
lattice. Since both terms contribute to the total kinetic
energy, the normalization $t^2+\lambda^2=1$ was imposed, and naturally
its square root is adopted as the unit of energy. With this
normalization, the kinetic energy turns out to be approximately
constant as $\lambda/t$ is varied \cite{meza} with all the remainder
parameters held fixed. The interacting part of the Hamiltonian
contains a ferromagnetic exchange coupling between conduction
electron spins ${\bf s}_l$ and localized magnetic moments ${\bf S}_l$,
with strength $J_{sd}$, and an exchange magnetic interaction between
localized magnetic moments with coupling $J$. This exchange term
favors a FM ($J<0$) or an AFM ($J>0$) order of the localized moments.

This system is schematically
shown in Fig.~\ref{fig1}(a)-(b). The plane of
localized magnetic moments (upper slab in both pictures) is
coupled by the exchange integral $J_{sd}$ to the layer where
conduction electrons are able to undergo conventional hopping and
Rashba type hopping (bottom slab).

The localized magnetic moments are assumed classical variables with
modulus equal to one, and hence they are described in spherical
coordinates by the angles $(\theta, \varphi)$. Notice that by
taking $|{\bf S}|=1$, the magnitude of the physical magnetic moments
has been absorbed in $J_{sd}$.

Model (\ref{ham1orb}) will be studied by exact diagonalization on
strips $L\times W$, where periodic boundary conditions are assumed 
in the longitudinal ($x$-axis) direction, and open boundary conditions
on the transversal ($y$-axis) direction. The electron filling, as in
all lattice models, is defined as the total number of electrons
divided by the total number of orbitals of the conducting layer, 
which for the present single-orbital model is equal to the number of
sites $N=L  W$. For all the sets of parameters considered, the 
condition of closed shell, that is, that all the degenerate 
single-electron eigenvalues up to the Fermi level were included, was 
verified in order to avoid spurious values of the physical properties
computed. In some cases, this condition was enforced by adding a small
twist $\Phi_x=10^{-7}$ in the boundary conditions along $x$.

The non coplanar spiral order of the magnetic moments is defined for 
the angle $\theta_l$ and a uniform azimuthal angle, $\varphi_l$, as:
\begin{eqnarray}
\theta_{x,y}={\bf k}_\theta \cdot (x, y)
\label{spiral}
\end{eqnarray}
with spiral momentum
${\bf k}_\theta=(k_{\theta,x},k_{\theta,y})=2\pi (m/L, m'/W)$,
$m,m'$ integers. In the following, the azimuthal angle will be
considered uniform, that is, $\varphi_l=\varphi$. Special cases are
the FM order, with $m=m'=0$, and the AFM order, with $m=L/2$, $m'=W/2$.

The system defined by the Hamiltonian given by Eq.~(\ref{ham1orb}) will
be studied at zero temperature and in linear response. For each set of
parameters, $\lambda/t$, $J_{sd}$, $J$, $W$, $L$, and $n$, and for each
pair of integers $(m,m')$, $m \in [0,L]$ and $m' \in [0,W]$, the total
energy is computed. The optimal spiral state for that set of parameters
is the one corresponding to the pair $(m,m')$ for which the minimum 
value of the total energy is obtained. All physical properties for
each set of parameters will be computed for the corresponding optimal
spiral momentum. Most of the calculations were performed on clusters
containing up to 8192 sites, although most of the results reported
below were obtained for $512\times W$ clusters.

The parameter $\lambda/t$ was varied in the interval $[0,1]$,
and  $J_{sd}$ was varied between 0 and 15. Since the effect of $J$
is somewhat trivial, in the following it will be set equal to zero.
For $J=0$, the FM order exists only for $\lambda=0$, and the AFM 
order appears in a finite region of the $\lambda/t$-$J_{sd}$ plane
at $n=0.5$, as discussed in the following section. For this density,
the FM or AFM orders exists for all the range of $\lambda/t$-$J_{sd}$
for $|J|\gtrapprox 1$. However, to compute physical properties for
these two orders is technically much simpler to fix the corresponding
values of $m,m'$, as mentioned above, while setting $J=0$.

All the transport properties studied below involve the charge current,
which is the sum of the spin-conserving current,
$J_{\sigma,\hat{\mu}}$, $\sigma=\uparrow,\downarrow$,
$\hat{\mu}=x, y$, which is the expectation value of the operator:
\begin{eqnarray}
\hat{j}_{\sigma,l,\hat{\mu}} = i t
(c_{l+\hat{\mu},\sigma}^\dagger c_{l,\sigma} - H. c.),
\label{curhop}
\end{eqnarray}
in units where the electron charge $e=1$, and of the spin-flipping
current, $J_{SO,\hat{\mu}}$ which is the
expectation value of the operator:
\begin{eqnarray}
\hat{j}_{SO,l,\hat{x}}&=& - i \lambda (c_{l+x\downarrow}^\dagger
   c_{l\uparrow}  -c_{l+x\uparrow}^\dagger c_{l\downarrow} - H. c.)
\nonumber  \\
\hat{j}_{SO,l,\hat{y}}&=& \lambda (c_{l+y\downarrow}^\dagger
   c_{l\uparrow} + c_{l+y\uparrow}^\dagger c_{l\downarrow} + H. c.)
\label{curso}
\end{eqnarray}

Other physical quantities involving also the transversal spin currents,
will be defined below.

\section{Phase diagram at quarter filling}
\label{phdiag}

Let us start by examining the phase diagram in the
$\lambda/t$-$J_{sd}$ plane at $n=0.5$, $J=0$. This phase diagram,
schematically shown in Fig.~\ref{fig1}(c), is approximately valid
for all strip widths, from the narrowest strip that could contain
Rashba helical currents and spin polarization across its section,
which corresponds to $W=2$ \cite{riera}, up to $W=32$, for which
results are virtually indistinguishable from those of $W=64$.

This diagram contains the main phases to be examined in this study.
At large $J_{sd}$, for all values of $\lambda/t$ in the interval
$(0,1]$, a spiral (SP) order of the localized magnetic moments,
with a spiral momentum $(k_{\theta,x},0)$, and $\varphi=0$ or $\pi$,
shown in Fig.~\ref{fig1}(a), is present. As $J_{sd}$ is reduced
below $J_{sd}\approx 6$, another interesting order appears, the
''staggered" spiral (s-SP) phase, with
$\bf{k}_\theta=(k_{\theta,x},\pi)$, $\varphi=0$ or $\pi$, shown in
Fig.~\ref{fig1}(b), which exists for $\lambda/t > 0$ up to a value
$(\lambda/t)^*$ where the magnetic slab enters into an AFM,
$(\pi,\pi)$, phase that in turn extends up to
$\lambda/t=1$. The boundary between the s-SP and AFM regions is
located at $\lambda/t \sim 0.65$ for $J_{sd}=2.5$, and at
$\lambda/t \sim 0.8$ for $J_{sd}=5$. This crossover between s-SP
and AFM phases, with a jump in the longitudinal spiral momentum from
$\approx \pi/2$ to $\pi$, is of first order since there are two 
minima in the energy as a function of $k_{\theta,x}$. For fixed
$\lambda/t$, the crossover between the s-SP and SP phases as
$J_{sd}$ is varied, is also first order. For $J_{sd}\lessapprox 1$
there are many competing phases depending strongly on the parameters
of the model.

\begin{figure}[t]
\begin{center}
\includegraphics[width=0.9\columnwidth,angle=0]{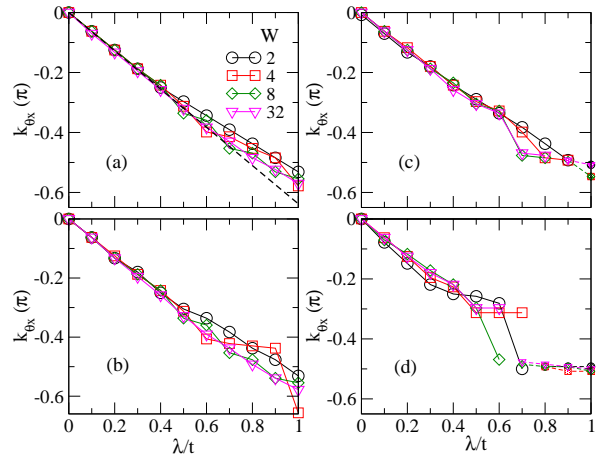}
\caption{(Color online) Variation of the
spiral momentum $k_{\theta,x}$ with $\lambda/t$ in the spiral (SP)
phase for (a) $J_{sd}=10$, and (b) $J_{sd}=7.5$. The dashed line is
a linear interpolation of the results for $W=32$ in $[0,0.6]$.
Variation of the spiral momentum $k_{\theta,x}$ in
the staggered spiral (s-SP phase) as a function of $\lambda/t$,
for (c) $J_{sd}=5$, and (d) $J_{sd}=2.5$. Symbols for various
strip widths $W$, are indicated on the plot. In (c) and (d) large
symbols and full lines correspond to spiral states that minimize
the Hamiltonian, while small symbols and dashed lines correspond
to spiral states that are excited states. Results for $n=0.5$.}
\label{fig2}
\end{center}
\end{figure}

An important feature in these spiral phases is the following. As it
can be observed in Fig.~\ref{fig2}(a), the spiral momentum along the
strip axis, $k_{\theta,x}$, decreases almost linearly from zero to
$\approx -\pi/2$ as $\lambda/t$ increases from zero to one, for
$J_{sd}=10$. This linear behavior is apparent from the interpolation
of the results for $W=32$ up to $\lambda/t \le 0.6$. However, for
larger values of $\lambda/t$, $k_{\theta,x}$ clearly starts to deviate
from that linear behavior. A similar behavior is shown in
Fig.~\ref{fig2}(b) for $J_{sd}=7.5$, also within the SP region.
Within the staggered spiral
phase, for $J_{sd}=5$, it can be also observed an almost linear
decrease of the longitudinal spiral momentum $k_{\theta,x}$, from zero
to $\approx \pi/2$ as $\lambda/t$ increases from zero to its maximum
value before entering in the AFM phase (Fig.~\ref{fig2}(c)). As said
above, this transition is of first order, and within the AFM phase,
the SP order corresponds to the first excited state within the subset
of states considered. The same behavior is observed for $J_{sd}=2.5$
except that in this case the AFM phase starts at a lower value of
$\lambda/t$ (Fig.~\ref{fig2}(d)). These results correspond to
$\varphi=0$. The same chirality of the reported spiral
states is recovered for $\varphi=\pi$ and reversing the sign of
$k_{\theta,x}$.

Since, as discussed in the Introduction, the spiral order of the
localized magnetic moments is driven by the conduction electrons,
it is expected that $J_{sd}$ will make this conducting-induced
spiral order on the localized magnetic slab to survive for larger
values of $\lambda/t$. This corresponds to the behavior shown in
Fig.~\ref{fig2}, where it can also be noticed that the spiral
momentum is roughly independent of $J_{sd}$ for a given value of
$\lambda/t$, as long as the spiral order exists. It should also be
emphasized that the relationship $k_{\theta,x} \sim \lambda/t$
remains valid except when approaching the value of
$k_{\theta,x} =\pi/2$, for the strip geometry here considered.
This departure of the linear behavior could be due to higher
order effects in $\lambda/t$, involving virtual processes through
$J_{sd}$, which were neglected in the first order calculation in
Ref.~\cite{kim2013}.

Notice also that 
in the absence of Rashba SO coupling, $\lambda=0$, and for large
$J_{sd}$, as in the conventional double-exchange model, and as
exemplified by manganites \cite{mangarep}, localized spins acquire a
FM, $(0,0)$ state, in order to favour the kinetic energy of conduction
electrons. For small values of $J_{sd}$, on the other hand, for
$\lambda=0$, the localized moments present a $(0,\pi)$ order.
Hence, although the variation of $k_{\theta,x}$ with $\lambda/t$
seems a result of the precession of single conducting electrons,
the presence of the SP, s-SP or AFM phases depends on the values of
$J_{sd}$, $\lambda/t$, and as it will be mentioned below, also on the
electron filling, through the full many-body nature of the system.

The location of the boundaries between different phases is mildly
dependent on the strip width $W$ in the proximity of quarter-filling.
For $n=0.25$ the phase diagram contains essentially the SP phase
for all values of $\lambda/t$ and $J_{sd} \gtrapprox 3$, with a
similar linear dependence of $k_{\theta,x}$ with $\lambda/t$ above
discussed for the $n=0.5$. The s-SP phase has disappeared. At
$J_{sd} = 2.5$, $W=4$, the localized moments have a 
$(\approx \pi/4,0)$ order for $\lambda=0$, and $k_{\theta,x}$
decreases by increasing $\lambda/t$ but with a smaller slope than
for the cases shown in Fig.~\ref{fig2}. Although this electron filling
was not exhaustively explored, the comparison with the results for
$n=0.5$, suggests that many-body effects determine the possible
magnetic phases of the system. This conclusion stems from the 
well-known behavior of Kondo lattice models, where an effective
interaction between the localized magnetic moments mediated by the
conduction electrons appears at an effective level. This effective
interaction makes the order of the magnetic layer to depend on the
filling of the conduction layer.

Generalized Kondo lattice models like the one here studied, present
a phase separated state close to half-filling in two dimensions in
the absence of RSOC, as it is well-known from studies in the context
of manganites \cite{mangarep}. The presence of phase separation has
also been discussed for the Rashba system at the LaAlO$_3$/SrTiO$_3$
interface \cite{bovenzi,caprara}. Previous calculations for model
(\ref{ham1orb}) on the square lattice, with $J_{sd}=10$, in the
whole range of $\lambda/t$ in $[0,1]$, have shown that phase
separation occurs close to half-filling, $n \gtrapprox 0.75$, and 
moreover, that actually it is suppressed by increasing
$\lambda/t$ \cite{meza}. The situation discussed in
 \cite{bovenzi,caprara} could be present for smaller $J_{sd}$,
particularly in the AFM region, but its precise determination 
is out of the scope of the present work.

Finally, it should be noticed that the reported spiral states are the
true ground state states as it results from Monte Carlo simulations
up to $\lambda/t \approx 0.5$. For larger SO couplings,
this phase is unstable towards various other orderings, one of them
is the already mentioned AFM for small $J_{sd}$, which is also the
true ground state. Moreover, Monte Carlo calculations show that for
$\lambda/t \gtrapprox 0.7$ and $J_{sd}$ above the AFM region, the
$(\approx \pi/2,0)$ spiral becomes unstable against other states
with a maximum of the magnetic structure factor near $(0, \pi/2)$.
These other states have an energy much lower than the one for the spiral
state with the spiral momentum equal to that maximum of the magnetic
structure factor, indicating that they are not spiral but a distinct,
so far unknown, order, probably a lattice of
skyrmions \cite{xli2014,yi2009}.

\section{Rashba helical currents}
\label{helicur}

The Rashba helical currents (RHC), with counter-propagating spin-up and
spin-down electron currents at each link at the lattice \cite{hamad}.
appear due to the presence of RSOC in the longitudinal directions on a
closed strip in equilibrium, that is in the absence of any external
electromagnetic field. Their presence can be inferred at an effective
level \cite{hamad} or by the mathematical structure of the RSOC 
Hamiltonian \cite{bovenzi,caprara}. A breaking of translation
invariance by adopting boundaries \cite{hamad}, or due to the presence
of impurities \cite{bovenzi} is necessary for the existence of the RHC.
The RHC are qualitative different to the spin currents (studied in the
next section) involved in the spin Hall effect that appear in the
transversal direction as a response to an injected charge current in
the longitudinal direction. These helical currents have been observed
in multiterminal devices (see references in \cite{hamad}).

In the following, results for the RHC correspond to the current of
spin-up electrons at each chain of the strip.

\begin{figure}[t]
\begin{center}
\includegraphics[width=0.9\columnwidth,angle=0]{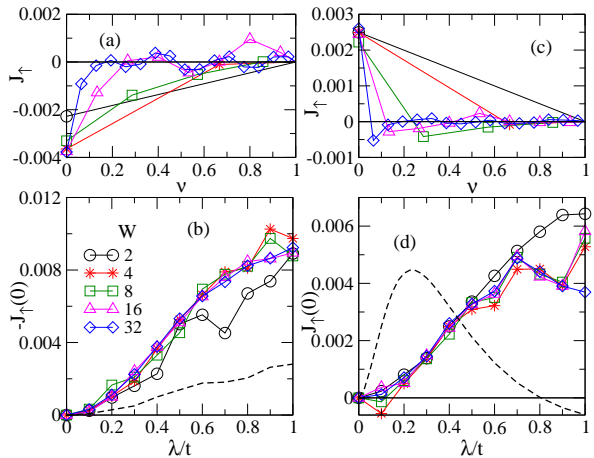}
\caption{(Color online) Current of spin-up electrons on each chain as
a function of the depth of the chain $\nu$ ($\nu=0$, edge, $\nu=1$,
center chain), for $\lambda/t=0.4$ and, (a) $J_{sd}=10$ (SP), and (b)
$J_{sd}=5$ (s-SP). Current of spin-up electrons on the edge chain,
$\nu=0$, as a function of $\lambda/t$ for (c) $J_{sd}=10$ and (d) 
$J_{sd}=5$.  Strip widths $W$ are indicated in the plot. In (c) and
(d) the edge currents for the fixed FM and AFM localized magnetic
slab are also included with dashed lines. In (d) the fixed AFM
results have been divided by $4$ in order to fit in the scale of the
plot. Electron filling $n=0.5$.
}
\label{fig3}
\end{center}
\end{figure}

In Figs.~\ref{fig3}(a) and (b), the current of spin-up electrons at
each chain on the planar strip, along the $x$-axis, is shown for the
SP ($J_{sd}=10$) and s-SP 
($J_{sd}=5$) regions respectively, as a function of the chain depth
$\nu$ ($\nu=0$, edge, $\nu=1$, center chain), for $\lambda/t=0.4$.
As expected, in both cases, the RHCs become concentrated at the
strip edges as $W$ is increased, and the decay of 
$J_\uparrow(\nu)$ with $\nu$ is faster for the staggered SP phase
than for the SP one, where oscillations can be still observed for
$W=32$.

To study the dependence of the RHC with $\lambda/t$, only the 
currents on the strip edge chain ($\nu=0$) will be considered.
Results for $J_\uparrow(0)$ as a function of $\lambda/t$ are shown
in Figs.~\ref{fig3}(c) and (d) for the SP ($J_{sd}=10$) and s-SP
($J_{sd}=5$) regions respectively, for various strip widths. Notice
that in Fig.~\ref{fig3}(c), SP region, the sign of the currents has
been changed in order to have a better comparison with the s-SP
case.

It can be observed that the dependence of $|J_\uparrow(0)|$ with 
$\lambda/t$ is similar for both the SP and s-SP states, particularly
for larger strip width where the results are also smoother, but the
RHC are larger for the SP case. There
is an approximately quadratic dependence for small $\lambda/t$,
as predicted in Ref.~\cite{hamad}, saturating as $\lambda/t$
approaches one. Notice an irregular behavior in the case of the 
s-SP (Fig.~\ref{fig3}(d)) for large $\lambda/t$, when the spiral
order becomes an excited state and the system enters in the AFM
region.

It is also interesting to compare the present results, where as
said above, the spiral order is driven by the SO coupling in the
conducting strip, to the two cases more considered in the previous 
literature \cite{riera2017} where a FM or an AFM order is fixed
by a large exchange coupling $|J|$ for all values of $\lambda/t$
and $J_{sd}$. By comparing the SP state with the fixed FM system,
one should notice that the RHC for the later (dashed line in
Fig.~\ref{fig3}(c)) is approximately three times smaller than for
the SP one for the same $W=32$ strip, that is, the SP order favours
the tendency to induce the RHC. On the other hand, by comparing the
s-SP state with the fixed AFM order, it is remarkable that the RHC
for the later (dashed line in Fig.~\ref{fig3}(d)) are much larger than
those for the staggered SP state for $\lambda/t \lessapprox 0.5$,
and of course much larger than the ones for the fixed FM case as 
originally noticed in Ref.~\cite{riera2017}.

It will be examined in the following section if any of these behaviors
translate into transport properties that are more conventionally
experimentally measured and more relevant for spintronics applications.

\section{Transport properties}
\label{perflux}

The zero temperature optical conductivity is defined as the real
part of the linear
response to the electric field and can be written as \cite{fye}:
\begin{align}
\sigma(\omega)&=D~\delta(\omega) + \sigma^{reg}(\omega) \label{optcond}\\
&=D~\delta(\omega)+ \frac{\pi}{L} \sum_{n \neq 0}
    \frac{| \langle \Psi_n | \hat{j}_x | \Psi_0 \rangle |^2}{E_n-E_0}
    \delta(\omega - (E_n-E_0))
\nonumber
\end{align}
where $| \Psi_n\rangle$ are the eigenstates of the total Hamiltonian
with energy $E_n$, $| \Psi_0 \rangle$ is the ground state, and the
paramagnetic current along the $x$-direction is defined in
terms of the currents defined in Eqs.~(\ref{curhop},\ref{curso}) as:
\begin{eqnarray}
\hat{j}_x&=& \hat{j}_{hop,x} + \hat{j}_{SO,x} \nonumber \\
\hat{j}_{hop,x}&=& \hat{j}_{\uparrow,x} + \hat{j}_{\downarrow,x}
 \nonumber \\
\hat{j}_{\sigma,x} &=& \sum_{l} \hat{j}_{\sigma,l,x}, 
     ~~~\sigma=\uparrow,\downarrow \nonumber \\
\hat{j}_{SO,x} &=&  \sum_{l} \hat{j}_{SO,l,x}
\label{current}
\end{eqnarray}
The Drude weight $D$ is calculated from the f-sum rule as:
\begin{eqnarray}
\frac{D}{2\pi} = -\frac{\langle H_{0,x} \rangle}{2L} - I_{reg}
\label{drude}
\end{eqnarray}
where $K_x \equiv -\langle H_{0,x} \rangle$ is the total kinetic
energy of electrons along the $x$-direction, and
\begin{eqnarray}
I_{reg} = \frac{1}{L} \sum_{n \neq 0}
\frac{|\langle \Psi_n | \hat{j}_x | \Psi_0 \rangle |^2}{E_n-E_0}
\label{intesreg}
\end{eqnarray}
is the integral over frequency of $\sigma_{reg}$ defined in 
Eq.~(\ref{optcond}). Clearly, $I_{reg}$, and hence $D$, will have
contributions from matrix elements 
$i_{\mu}=| \langle \Psi_n | \hat{j}_{\mu,x} | \Psi_0 \rangle |^2$,
($\mu=$~hop, SO), and 
$i_{crossed}=2 Re[\langle \Psi_n | \hat{j}_{hop,x} | \Psi_0 \rangle
\langle \Psi_0 | \hat{j}_{SO,x} | \Psi_n \rangle ]$.

\begin{figure}[t]
\begin{center}
\includegraphics[width=0.9\columnwidth,angle=0]{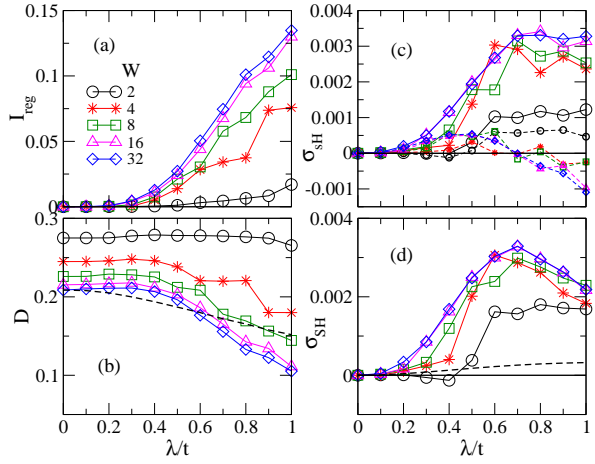}
\caption{(Color online) (a) Integral of the regular part of the
optical conductivity, (b) Drude weight, (c) $\langle h_x h_y\rangle$
(full lines) and $\langle SO_x h_y\rangle$ (dashed
lines) contributions to the spin Hall conductivity (see text for 
definitions), and (d) total spin Hall conductivity as a function of
$\lambda/t$ for various strip widths $W$ indicated on the plot.
In (b) and (d) the corresponding values for the fixed FM state,
$W=32$, were added for comparison. Spiral phase, $J_{sd}=10$.}
\label{fig4}
\end{center}
\end{figure}

The spin Hall conductivity is the main quantity describing the 
spin Hall effect because it involves spin currents appearing in the
transversal direction when a charge current is applied to a conductor
in the longitudinal direction. 
The spin Hall conductivity $\sigma_{sH}$ is defined as the
$\omega=0$ limit of the spin-charge transversal response function
given by the Kubo formula, at zero temperature \cite{rashba04,sinova04}:
\begin{align}
\sigma^{sc}_{xy}(\omega) &=& -i \frac{t}{\pi \lambda} \sum_{n}
      \sum_{m} 
      \frac{\langle \Psi_n | \hat{j}^s_y | \Psi_m \rangle
     \langle \Psi_m | \hat{j}_x | \Psi_n \rangle}{[(E_n-E_m)^2-\omega^2]}
\label{sHcond}
\end{align}
where $j^s_y$ is the spin current along the $y$-direction. In the
first sum, the summation is performed only over states with energies 
$E_n$ larger than the Fermi energy $E_F$, and in the second sum only
over states with energies $E_m < E_F$. The spin current operator at
each bond connecting sites $l$ and $l+\hat{v}$, where $\hat{v}$ is
the unit vector along the $x$ or $y$ directions, follows from the
{\em local} spin conservation operator equation in the absence of
external torques: $\sum_{\hat{v}} \hat{j}^s_{l,\hat{v}} +
\partial \hat{S}^z_l/\partial \tau=0$ ($\tau$ is the time). For the
total $\hat{j}^s_y$ the following
expression is obtained \cite{riera2017}:
\begin{eqnarray}
\hat{j}^s_y&=&\hat{j}^s_{hop,y} + \hat{j}^s_{SO,y},
\nonumber  \\
\hat{j}^s_{hop,y}&=&\frac{1}{2}(\hat{j}_{\uparrow,y} - 
                    \hat{j}_{\downarrow,y}),
\nonumber  \\
\hat{j}^s_{SO,y}&=& - \frac{\lambda}{2} \sum_{l}
   (c_{l+y\downarrow}^\dagger c_{l\uparrow} -
    c_{l+y\uparrow}^\dagger c_{l\downarrow} + H. c.)
\label{spincury}
\end{eqnarray}
This expression of $\hat{j}^s_y$ is the second quantized equivalent to
the one formulated in first quantization and using a parabolic kinetic
energy that was considered in previous calculations of the spin Hall 
conductivity \cite{rashba04,sinova04}.

By replacing the longitudinal charge currents and the transversal
spin currents into Eq.~(\ref{sHcond}), in the same way as it was done
for the optical conductivity, the contributions to the double sum can
be classified according to the conserving or non-conserving currents
involved. For example, there is a contribution from terms
$\langle \Psi_n | \hat{j}^s_{hop,y} | \Psi_m \rangle
\langle \Psi_m | \hat{j}_{hop,x} | \Psi_n \rangle$,
($\langle h_x h_y\rangle$ for short), and another contribution from
$\langle \Psi_n | \hat{j}^s_{hop,y} | \Psi_m \rangle
\langle \Psi_m | \hat{j}_{SO,x} | \Psi_n \rangle$,
($\langle SO_x h_y\rangle$ for short).

\begin{figure}[t]
\begin{center}
\includegraphics[width=0.9\columnwidth,angle=0]{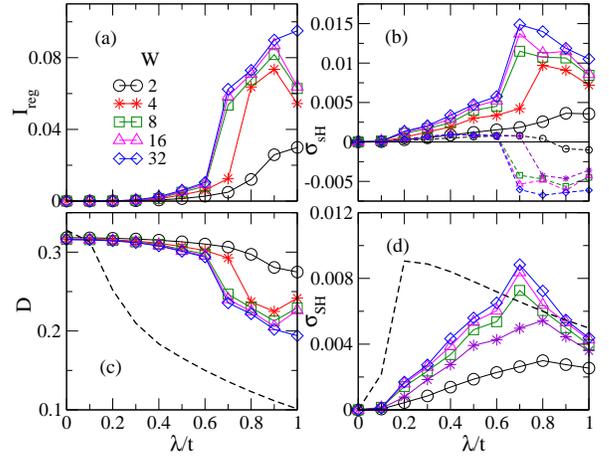}
\caption{(Color online) (a) Integral of the regular part of the
optical conductivity, (b) Drude weight, (c) $\langle h_x h_y\rangle$
(full lines) and $\langle SO_x h_y\rangle$ (dashed
lines) contributions to the spin Hall conductivity (see text for 
definitions), and (d) total spin Hall conductivity as a function of 
$\lambda/t$ for various strip widths $W$ indicated on the plot.
In (b) $D$ for the fixed AFM state, $W=32$, was added for comparison
multiplied by 2.
In (d) $\sigma_{sH}$ for the fixed AFM state, $W=32$,
was added with the sign changed and divided by 2.
Staggered spiral phase, $J_{sd}=5$.}
\label{fig5}
\end{center}
\end{figure}

In Fig.~\ref{fig4}(a), the integral of the regular part of the 
longitudinal optical conductivity, $I_{reg}$, defined
in Eq.~(\ref{intesreg}), is shown for various strip widths as a 
function of $\lambda/t$ for the spiral state with $J_{sd}=10$.
A strong increase of $I_{reg}$ can be observed as $W$ is
increased and for large $\lambda/t$. Both matrix elements $i_{hop}$
and $i_{SO}$ contribute to $I_{reg}$, with a slightly larger 
contribution from $i_{hop}$, and the contribution from the crossed
matrix elements changes sign from negative to positive at $W=8$.
The Drude peak, shown in Fig.~\ref{fig4}(b), has two noticeable
behaviors. In the first place, $D$ decreases as the strip width
increases in the whole range of $\lambda/t$. This is clearly due
to a reduction of the kinetic energy with increasing $W$, since
$I_{reg}$ is negligible for small $\lambda/t$. For
large $W$ and $\lambda/t \gtrapprox 0.4$, a further decrease in
$D$ can be observed, in this case due to the increase in 
$I_{reg}$ shown in Fig.~\ref{fig4}(a). The Drude peak for the
fixed FM state \cite{riera2017}, for $W=32$, is close to the one for
the SP state for small $\lambda/t$, but it is clearly larger for
large $\lambda/t$, indicating the expected favouring of transport
by the FM state. 

Fig.~\ref{fig4}(c) shows the two nonzero contributions to the spin
Hall conductivity $\sigma_{sH}$ for various strip widths as a
function of $\lambda/t$ for the SP state. It can be seen that
the main contribution, $\langle h_x h_y\rangle$, increases with 
both $W$ and $\lambda/t$, while the $\langle SO_x h_y\rangle$
contribution decreases with $W$ and it even changes sign for 
large $\lambda/t$. The total $\sigma_{sH}$ is shown in 
Fig.~\ref{fig4}(d), showing a clear enhance with increasing $W$,
saturating at $W=32$. The spin Hall conductivity in the Rashba strip
with $W=32$ coupled to a FM layer was also added for comparison. For
the FM state, $\sigma_{sH}$, is about one order of magnitude smaller
than the one of the SP order, as it can be seen in Fig.~\ref{fig4}(d).
It should also be noticed that $\sigma_{sH}$ for the fixed FM state
is entirely due to the transitions $\langle SO_x h_y\rangle$
\cite{riera2017}.

Let us now examine transport properties for the staggered spiral
state. In the following, in the whole interval$0\le \lambda/t \le 1$
only results for the s-SP state will be reported even when it is an
excited state and the true ground state is the AFM state, as
discussed in Section~\ref{phdiag}.

In Fig.~\ref{fig5}(a), $I_{reg}$ is shown for various strip widths as
a function of $\lambda/t$ for $J_{sd}=5$. By comparing these results
with those for the SP phase shown in Fig.~\ref{fig4}(a) it is clear
in the first place that $I_{reg}$ is smaller for the s-SP state, for
the same values of $\lambda/t$ and $W$. In the second place, while in
the SP case, for a given $W$, $I_{reg}$ has a smooth behavior with
$\lambda/t$, particularly for wider strips, in the staggered case
there is a clear jump at $\lambda/t=(\lambda/t)^*$
at which the s-SP state becomes an excited state.
As in the SP case, the contribution from the matrix
elements $i_{hop}$ is slightly larger than the one from $i_{SO}$,
but for the staggered case the contribution from the crossed
matrix elements are negative for all $\lambda/t$ and $W$.
This behavior is translated to the Drude weight as shown in
Fig.~\ref{fig5}(b), where, for each $W$, a clear jump is observed
at the same values of $\lambda/t$ where a jump appears in 
$I_{reg}$. Overall, the Drude weight is larger for the staggered
case, where it is also absent the suppression of the kinetic energy
term pointed out regarding Fig.~\ref{fig4}(b). This is partially due
to a smaller value of the conducting-magnetic slabs, $J_{sd}$ adopted
for the staggered case. As expected, the Drude peak of the fixed AFM
state, also included in Fig.~\ref{fig5}(b) for $W=32$, is strongly
suppressed as $\lambda/t$ increases, which is opposed to the behavior
above discussed for the FM state. 

With respect to the spin Hall conductivity, it can be observed in 
Fig.~\ref{fig5}(c) that, similarly to what was noticed for the SP
case, in the s-SP state the main contribution comes from the terms
$\langle h_x h_y\rangle$. Both contributions suffer a jump again
at $(\lambda/t)^*$.  The $\langle h_x h_y\rangle$ becomes larger
with both $\lambda/t$ and $W$, while the $\langle SO_x h_y\rangle$
contribution becomes negative and increases in absolute value
with both $\lambda/t$ and $W$. As a result, the total
$\sigma_{sH}$, shown in Fig.~\ref{fig5}(d), increases with $W$
converging at $W \approx 32$, and reaching a maximum for each $W$ at
$(\lambda/t)^*$. Another
important conclusion is that $\sigma_{sH}$ for the staggered spiral
phase is about a factor of 3 larger than the one for the spiral
phase shown in Fig.~\ref{fig4}(d). The values of $\sigma_{sH}$ for
the fixed AFM state, $W=32$, \cite{riera2017} were also included in
Fig.~\ref{fig5}(d) for the sake of comparison. Notice that these
values were divided by 2, and their sign was changed in order to
fit into the plot scale. As for the fixed FM coupled magnetic
slab, $\sigma_{sH}$ for the AFM case is entirely due to the
$\langle SO_x h_y\rangle$ contribution. Notice that $\sigma_{sH}$
for the fixed AFM coupled slab is still much larger than the one for
the staggered spiral phase.

Finally, it is also interesting to discuss the intra- and inter-band
contributions to $\sigma_{sH}$, which correspond to transitions
between single-particle states with the same (opposite) chirality.
The chirality of each single-particle state
is defined by the sign of the $y$-component of the electron spin
averaged on that single-particle state. As discussed in 
Ref.~\cite{riera2017}, this definition appears as a natural extension
to strips of the corresponding concepts used for infinite 2D systems
\cite{rashba04,sinova04,ganichev}. Notice also that in the literature
alternative definitions have been used \cite{hli2015}.

\begin{figure}[t]
\begin{center}
\includegraphics[width=0.9\columnwidth,angle=0]{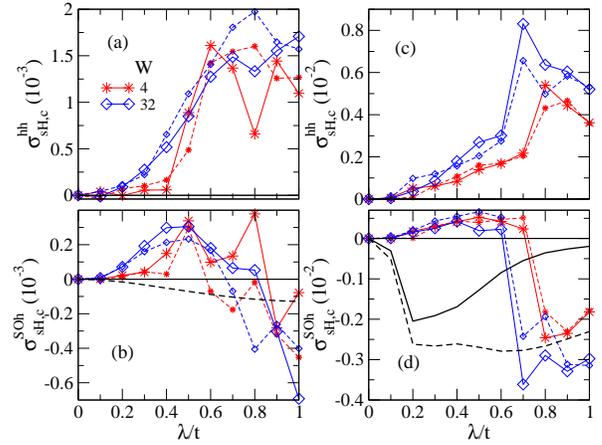}
\caption{(Color online) Intra- (full lines) and inter-band (dashed
lines) contributions to (a,c) $\langle h_x h_y\rangle$ and (b,d)
$\langle SO_x h_y\rangle$ terms of the spin Hall conductivity
as a function of $\lambda/t$ for various strip widths $W$
indicated on the plot.
(a,b) correspond to the SP state, $J_{sd}=10$, and (c,d) to the
s-SP state, $J_{sd}=5$. In (b) results for the interband contribution
to $\sigma_{sH}$ for the fixed FM state (thick dashed line) were 
added (multiplied by 4). In (d) results for the intra- (thick full
line) and interband (thick dashed line) contributions to 
$\sigma_{sH}$ for the fixed AFM state were added (divided by 2). 
}
\label{fig6}
\end{center}
\end{figure}

Results for both types of contributions to the spin Hall conductivity,
$\langle h_x h_y\rangle$ and $\langle SO_x h_y\rangle$, for the spiral
and staggered spiral states, are shown in Figs.~\ref{fig6}(a,c) and
Figs.~\ref{fig6}(b,d) respectively. For both types of spirals, and for
both types of contributions to $\sigma_{sH}$, a first conclusion is
that there is approximately the same contribution from both inter
and intra band transitions. For the spiral state, this behavior is
different than for the fixed FM order where only interband transitions
contribute. On the other hand, for the staggered spiral state, the 
behavior is similar to the one for the fixed AFM order where both 
types of processes are present although the interband ones are
dominant. A second conclusion is that, consistently with the results
presented so far, all the contributions have a rather smooth behavior
except when the system starts to deviate from the longitudinal spiral
order, which occurs for large $\lambda/t$. This change of behavior is
more clear for the $\langle SO_x h_y\rangle$ contribution, and even 
more notorious for the staggered spiral state when it becomes an 
excited state (Figs.~\ref{fig6}(d)).

To end this Section, let us briefly discuss the possibility of a spin
current along the longitudinal direction. This current has also two
components analogous to the expressions for $\hat{j}^s_y$ given by
Eq. (\ref{spincury}). The
hopping part, $\hat{j}^s_{hop,x}$ is conventionally termed the spin 
polarized longitudinal current. By replacing
$\hat{j}^s_x$ for $\hat{j}^s_y$ in the expression of the
spin Hall conductivity Eq.~(\ref{sHcond}), an analogous Kubo formula
at zero frequency for the spin polarized conductivity, $\sigma_{sp}$,
would be obtained. Let us call $\sigma_{sp,hh}$ the contribution to
$\sigma_{sp}$ from the hopping charge current and $\hat{j}^s_{hop,x}$.
It was shown that $\sigma_{sp,hh}$, as well as the anomalous Hall 
conductivity, $\sigma_{AH}$ \cite{AHERMP}, are equal to zero for the
pure Rashba
model \cite{inoue2003}. This topic was recently further discussed in
\cite{amin2016}. Preliminary calculations show that not only
$\sigma_{sp,hh}$ but the full $\sigma_{sp}$, as well as the anomalous
Hall conductivity, vanish for the spiral, staggered-spiral and AFM
states \cite{rieraprep}. For the FM state, $\sigma_{sp,hh}=0$, but
$\sigma_{sp}$ as well as $\sigma_{AH}$ are different for zero, for
the later, in agreement with \cite{onoda2008}.

\section{Conclusions}
\label{conclusions}

The first conclusion of this work is that when the magnetic exchange
between the magnetic moments is negligible, at quarter filling, a spiral
(SP) order of the localized magnetic moments with transversal momentum
equal to zero, that is, uniform across the strip section, exists for
interlayer exchange
coupling $J_{sd} \gtrapprox 5$, while another spiral order with
transversal momentum equal to $\pi$, that is, staggered across the
strip section, dominates for $J_{sd} \lessapprox 5$. In addition, this
s-SP order is unstable towards an antiferromagnetic phase for
$\lambda/t$ greater than a $J_{sd}$-dependent value. Both SP/s-SP and
s-SP/AFM crossovers are of first order.

The second conclusion is that both spiral phases have an almost
linear dependence of the spiral longitudinal momentum with $\lambda/t$
as long as this momentum is smaller than $\approx \pi/2$. Since this
linear behavior is essentially driven by the precession of independent
electrons moving on the conducting slab, one could speculate that for
large $\lambda/t$, effective interactions mediated by $J_{sd}$ lead
to a loss of coherence of the electron gas, particularly for the small
wavelengths corresponding to the longitudinal spiral momentum
approaching $\pi/2$. In the SP order, there is then a noticeable
departure from the linear behavior, and in the s-SP region, the whole
spiral is finally replaced by an AFM order. In principle, these spiral
states could be detected by neutron scattering techniques. However,
since the magnetic layer should be a thin film, and in addition, be
part of an heterostructure, there are other techniques that may be
more appropriate. For example, there is a recently devised technique
to detect spiral states on thin films using quantum sensors
that can achieve resolution of a few nanometers \cite{gross}.

As a summary of transport properties, including the Rashba helical
currents, it is clear than in general the SP and s-SP states 
interpolate between the behaviors previously observed for the fixed
FM and AFM orders. Still, there are considerable differences in the
amplitudes that these properties have in the SP and s-SP regions,
particularly a near factor of two in the spin Hall conductivities.
In addition, the momentum or the period of the spiral state can be
controlled by external electric fields through the RSOC leading to
a multiferroic behavior, and at the same time, the different responses
could also be exploited for spintronic applications. These results for
transport properties can be experimentally verified by the usual
techniques employed for studying the spin Hall effect, as reviewed
in \cite{sinovaRMP}.

A key ingredient of a device that could take advantage of the
present results, and that indicates a departure from the devices 
containing FM or AFM layers studied so far, is to have a very low
exchange coupling $J$ between the localized magnetic moments. These
virtually noninteracting magnetic moments could be created in the
first place by depositing magnetic Fe or CO atoms or nanoparticles
on top of a conducting Rashba layer. A second possibility is to employ
a ferromagnetic semiconductor for the magnetic coupled layer, at a
temperature above its Curie temperature, which for this class of
materials is very low, within a setup similar to that of
\cite{miron10}. For this effect to be robust, the magnetic slab
should be a thin film.

To some extent, a variation of $J_{sd}$ could also be achieved for
a single device by the
application of an external magnetic field perpendicular
to the planes, although certainly this would not be practical. On the
other hand, it is possible to set an electric field modulation of 
$J_{sd}$, which was intensively studied in the context of magnetic
storage devices \cite{ando2016}. In particular, the mechanism of
electric field modulation of the magnetic anisotropy consists in
changing the electron occupancy at d-orbitals in coupled 3d-transition
metal slabs by shifting the Fermi level and/or modifying the
electronic structure close to the Fermi level. Since $J_{sd}$ contains
the amplitude of the magnetic moments, it could acquire large values.
On the other hand, Since $J_{sd}$ is an effective coupling between the
conducting and the magnetic layers, it can be made arbitrarily small
by interposing layers of nonmagnetic insulating material between the
magnetic moments and the Rashba conducting layer as in 
\cite{manchon08}.

It is also worth to notice that the spiral state implies a breaking of
translational invariance of the system along the longitudinal direction
and hence a modulation at an effective level of the RSOC and hopping
couplings. In this sense, it would be interesting to investigate if
the presently studied system shows a spin Hall conductivity that 
remains robust against disorder as in a recently proposed model with
modulated RSOC \cite{seibold}.

Further study at finite
temperature and in out-of-equilibrium regimes would be necessary to
assess the usefulness of the various features here reported.

\section{Acknowledgments}
Useful discussions with A. Greco, C. Gazza, I. Hamad, and G. Meza,
are gratefully acknowledged.
The author is partially supported by the Consejo Nacional de
Investigaciones Cient\'ificas y T\'ecnicas (CONICET) of Argentina
through grant PIP No. 11220120100389CO.

\end{document}